\def\spose#1{\hbox to 0pt{#1\hss}}

\def\multleft#1{\hbox to size{\vbox {\halign {\lft{##}\cr #1}}\hfill}\par}
\def\multright#1{\hbox to size{\vbox {\halign {\rt{##}\cr #1}}\hfill}\par}

\def\today{\ifcase\month\or January\or February\or March\or April\or May\or
      June\or July\or August\or September\or October\or November\or December\fi
      \space\number\day, \number\year}
\def\s{\hbox{\phantom{5}}}	

\def\cm{{\rm\thinspace cm}}

\def\erg{{\rm\thinspace erg}}

\def\microJy{{\rm\thinspace $\mu$Jy}}

\def\K{{\rm\thinspace K}}
\def\keV{{\rm\thinspace keV}}
\def\km{{\rm\thinspace km}}
\def\kpc{{\rm\thinspace kpc}}
\def\Lsun{\hbox{$\rm\thinspace L_{\odot}$}}
\def\m{{\rm\thinspace m}}
\def\Mpc{{\rm\thinspace Mpc}}
\def\Msun{\hbox{$\rm\thinspace M_{\odot}$}}

\def\s{{\rm\thinspace s}}

\def\yr{{\rm\thinspace yr}}
\def\sr{{\rm\thinspace sr}}

\def\ergps{\hbox{$\erg\s^{-1}\,$}}

\def\kmps{\hbox{$\km\s^{-1}\,$}}

\def\Msunpyr{\hbox{$\Msun\yr^{-1}\,$}}

\def\psqcm{\hbox{$\cm^{-2}\,$}}

\def\kmpspMpc{\hbox{$\kmps\Mpc^{-1}$}}

\def\H2{\hbox{H$_{2}$}}

\documentclass[usegraphicx]{mn2e}

\usepackage{times}
\usepackage{amssymb}
\usepackage{eso-pic}
\usepackage[usenames]{color}

\include{defn}

\begin{document}
\hsize=6truein

\title[A semi-empirical simulation of the extragalactic radio continuum sky]{A semi-empirical simulation of the extragalactic radio continuum sky for next generation radio telescopes}
\author[R.J.~Wilman et al.]
{\parbox[]{6.in} {R.J.~Wilman$^{1}$, L.~Miller$^{1}$, M.J.~Jarvis$^{2}$, T.~Mauch$^{1}$,  F.~Levrier$^{1}$, F.B.~Abdalla$^{3}$, S.~Rawlings$^{1}$, H.-R.~Kl\"{o}ckner$^{1}$, D.~Obreschkow$^{1}$, D.~Olteanu$^{4}$, S.~Young$^{4}$ \\ \\
\footnotesize
1. Oxford Astrophysics, Denys Wilkinson Building, Keble Rd, Oxford, OX1 3RH \\ 
2. Centre for Astrophysics, Science \& Technology Research Institute, University of Hertfordshire, Hatfield, AL10 9AB \\
3. Department of Physics \& Astronomy, University College London, Gower St., London, WC1E 6BT \\
4. Oxford e-Research Centre, 7 Keble Road, Oxford, OX1 3QG \\ }}
\maketitle

\begin{abstract}
We have developed a semi-empirical simulation of the extragalactic radio continuum sky suitable for aiding the design of next generation radio interferometers such as the Square Kilometer Array (SKA). The emphasis is on modelling the large-scale cosmological distribution of radio sources rather than the internal structure of individual galaxies. Here we provide a description of the simulation to accompany the online release of a catalogue of $\simeq$ 320 million simulated radio sources. The simulation covers a sky area of $20 \times 20$~deg$^{2}$ -- a plausible upper limit to the instaneous field of view attainable with future (e.g. SKA) aperture array technologies --  out to a cosmological redshift of $z=20$, and down to flux density limits of 10 nJy at 151, 610~MHz, 1.4, 4.86 and 18~GHz. Five distinct source types are included: radio-quiet AGN, radio-loud AGN of the FRI and FRII structural classes, and star-forming galaxies, the latter split into populations of quiescent and starbursting galaxies.

In our semi-empirical approach, the simulated sources are drawn from observed (or extrapolated) luminosity functions and grafted onto an underlying dark 
matter density field with biases which reflect their measured large-scale clustering. A numerical Press-Schechter-style filtering of the
density field is used to identify and populate clusters of galaxies. For economy of output, radio source structures are 
constructed from point source and elliptical sub-components, and for FRI and FRII sources an orientation-based unification and 
beaming model is used to partition flux between the core and extended lobes and hotspots. The extensive simulation output gives 
users the flexibility to post-process the catalogues to achieve more complete agreement with observational data in the years ahead.
The ultimate aim is for the `idealised skies' generated by this simulation and associated post-processing to be fed to telescope simulators to optimise the design of the SKA itself. 
\end{abstract}


\section{INTRODUCTION}
The {\em Square Kilometer Array} (SKA) is the next generation radio telescope facility for the 21st century. Although it is 
still in the design stage and unlikely to be fully operational before 2020, various pathfinder experiments will enter 
scientific service around 2010; one of these pathfinders will probably form the nucleus of the eventual SKA and over the next 
decade grow in sensitivity from $<1$ to 100 per cent SKA. The scientific goals of the SKA have been described at length 
elsewhere (see e.g. Carilli \& Rawlings~2004), but it is now clear that a new generation of more sophisticated science simulations 
is needed to optimise the 
design of the new telescopes 
and their observing programmes for the efficient realisation of these goals. With this in mind, we present here a new semi-empirical
simulation of the extragalactic radio continuum sky. The simulation is part of a suite of simulations developed under the European SKA Design Study (SKADS) initiative and referred to collectively as {\em SKADS Simulated Skies ($S^{3}$)}. It is one of two simulations
aimed at simulating the extragalactic continuum and line-emitting radio sky, which together offer distinct 
yet complementary approaches to modelling the radio sky. The first approach, used by the simulations described in this paper, may be best 
described as `semi-empirical', in the sense that the simulated sources are generated by sampling the observed (or extrapolated) 
radio continuum luminosity functions. The second approach, pursued by Obreschkow et al.~(2008), is dubbed `semi-analytical' 
as it ascribes gas properties, star formation and black hole accretion rates to galaxy haloes identified in an N-body simulation.
Given their different emphases, the simulations are also commonly referred to as the `continuum' (semi-empirical) and `HI' (semi-analytical) simulations and, where appropriate, synergies between them will be highlighted. Limitations in state-of-the-art N-body 
simulations mean that the semi-empirical simulations cover $\sim 100$~times more sky area than the semi-analytical ones.

The purpose of the present paper is to describe the essential methodology and basic ingredients of the semi-empirical 
simulation (section 2); to demonstrate that its output satisfies some basic tests, such as reproducing the radio source counts, local
luminosity function and angular clustering (section 3); to highlight some known deficiencies of the simulations (section 4). 
Simulation data products (source catalogues and images) can be accessed through a database web-interface (http://s-cubed.physics.ox.ac.uk) where full details of data format can be found.

\section{DESCRIPTION OF THE SIMULATION}

There have been several previous attempts to model the radio source counts at the faint flux levels appropriate for the SKA and 
to produce simulated source catalogues, e.g. Hopkins et al.~(2000), Windhorst~(2003), Jackson~(2004), Jarvis \& Rawlings~(2004). 
Our new simulation builds on these efforts and, like them, is empirically based in the sense that it draws sources at random from 
the observed (or suitably extrapolated) radio continuum luminosity functions for the extragalactic populations of interest. 
In a major advance over these works, however, we have attempted to model the spatial clustering of the sources instead of simply
placing them on the sky with a uniform random distribution. This is an important feature because large scale
structure measurements have emerged as a key SKA goal, stemming from the possibility to constrain the properties of dark 
energy through the detection of baryon acoustic oscillations (BAO) (e.g. Abdalla \& Rawlings~2005). The spatial clustering has 
been incorporated on large scales ($\gg5 h^{-1}$\Mpc) by biasing the galaxy populations relative to an underlying dark matter density field which is evolved 
under linear theory. On smaller scales, we have also used the underlying dark matter density field to identify and populate 
clusters of galaxies.

Another major impetus for the development of the new simulations has been the desire to generate an extensive suite of output 
catalogues which can be stored as a database and queried via a public web interface. The aim is to use such a tool 
to generate `idealised skies' which can then be fed to telescope simulators for the production and 
analysis of artifical datasets. Such a process is essential for optimising the design of the telescopes and their observing
programmes.

\subsection{Design requirements of the simulation}

The semi-empirical simulation has been designed with the key SKA science goals in galaxy evolution and large-scale structure in 
mind. The basic parameters are as follows:

\noindent
$\bullet$ {\em Sky Area:} $20 \times 20$ deg$^{2}$, approximately matching the largest plausible instantaneous field of view of a future SKA aperture
array. The latter will be used to perform HI redshift surveys out to $z \simeq 1.5$ over a significant fraction of the sky, in 
order to measure the galaxy power spectrum and the evolution of dark energy via baryon acoustic oscillations (BAO) (see e.g. 
Abdalla \& Rawlings~2005).

\noindent
$\bullet$ {\em Maximum redshift:} $z = 20$. Probing the ``dark ages'' in HI and radio continuum is another key SKA goal (see e.g. Carilli 2008), and the redshift limit of the simulation has accordingly been set to include the reionization epoch. 

\noindent
$\bullet$ {\em Flux density limits:} 10~nJy at the catalogue frequencies of 151~MHz, 610~MHz, 1.4~GHz, 4.86~GHz and 18~GHz, 
spanning the expected operational frequency range of the SKA. At 1.4~GHz, 10~nJy is of the order of the expected sensitivity
for a 100h observation with the nominal SKA sensitivity ($A/T_{\rm{sys}}=20000$~\m$^{2}$ \K$^{-1}$, 350~MHz bandwidth). The flux limits are
applied such that if any structural sub-component of a galaxy (i.e. core, lobes or hotspots in the case of a radio-loud AGN) 
has a flux density exceeding this limit at one or more of these frequencies, all sub-components of the galaxy are included in the 
output catalogue, regardless of whether they are all above the flux limit. In this paper, we neglect any intrinsic or induced 
polarization properties of the radio emission, but we expect the first major revision of the catalogue to include such information.

\noindent
$\bullet$ {\em Source populations:} In addition to the classical double-lobed radio-loud AGN of the FRI and FRII classes, 
the simulation will also incorporate radio emission from `radio-quiet' AGN and star-forming galaxies. The latter population 
comprises relatively quiescent or `normal' late-type galaxies, as well as more luminous and compact starburst galaxies. Together 
these two populations are expected to dominate the radio source counts below a few 100\microJy~at 1.4~GHz.

\subsection{The theoretical framework}
\begin{figure}
\includegraphics[width=0.48\textwidth,angle=0]{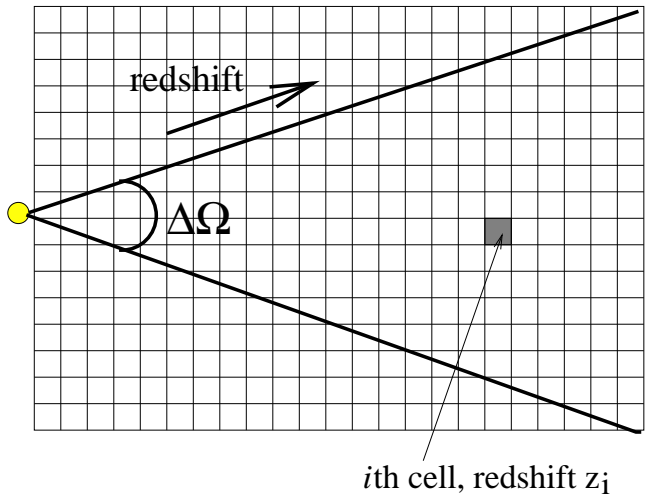}
\caption{\normalsize A 2-dimensional illustration of the basic geometry of the simulation. As discussed in section 2.2, the
simulation consists in looping over all cells (each of which is $5 h^{-1}$\Mpc~in size) contained within a cone of solid angle 
$\Delta \Omega$, at the apex of which sits an imaginary observer. At the position of the $i$th such cell (shaded), the co-moving 
distance to the observer is inverted to yield the redshift, $z_{\rm{i}}$, of the cell.}
\label{fig:grid}
\end{figure}

The starting point for the simulation is a present-day ($z=0$) dark matter density field, $\delta \rho/\rho$, defined on a 
cuboid grid of 5$h^{-1}$\Mpc~comoving cells with the overall array having dimensions $550 \times 550 \times 1550$~cells (the long axis defines 
the direction of increasing redshift). An imaginary observer is situated at the centre of one face of the grid and 
looks out into the simulation volume over a solid angle $\Delta \Omega$. The simulation consists in looping over all cells 
contained within $\Delta \Omega$ (see illustration in Fig.~\ref{fig:grid}). At the position of the $i$th cell, the comoving 
distance to the observer is inverted to yield an observed redshift, $z_{\rm{i}}$, for the assumed cosmological model. For each source type (denoted by suffix $j$), the observed-frame spectral energy 
distribution is compared with the flux density limits of the simulation to derive a minimum luminosity, 
$L^{\rm{i}}_{\rm{min,j}}$, for sources to be included in the output catalogues. The luminosity function is integrated down to
this limit to yield the space density, $\phi(L>L^{\rm{i}}_{\rm{min,j}},z_{\rm{i}})$. In the absence of any large-scale
structure, the mean number of sources expected in the cell would be simply $n_{\rm{exp,j}}^{\rm{i}} = \phi(L>L^{\rm{i}}_{\rm{min,j}},z_{\rm{i}}) \Delta V$, where $\Delta V$ is the co-moving cell volume. To introduce large-scale structure into the distribution of the sources, we modify this as follows:

\begin{equation}
n_{\rm{exp,j}}^{\rm{i}} = A e^{b(z_{\rm{i}})G(z_{\rm{i}}) \delta \rho/\rho} \phi(L>L^{\rm{i}}_{\rm{min,j}},z_{\rm{i}}) \Delta V,
\end{equation} 
where G(z) is the linear growth factor for fluctuations in the underlying dark matter density field, b(z) is the redshift-dependent 
bias of the population of interest, and $A$ is a normalisation factor which ensures agreement with the luminosity function when averaged over the largest scales. For a gaussian density field, it is easily shown that $A=e^{-b(z)^2 G(z)^2 \sigma^2/2}$, where $\sigma^2$ is
the cell variance of the density field. Simulated sources are generated by Poisson-sampling the luminosity function at 
$L>L^{\rm{i}}_{\rm{min,j}}$ with mean $n_{\rm{exp,j}}^{\rm{i}}$.

The formalism of eqn.~(1) is a somewhat adhoc, non-linear bias prescription which has the effect of amplifying the source density 
in over-dense regions and depressing it in under-dense regions. When the exponent is small it reduces to a linear bias prescription, i.e. $\delta n/n \simeq  b(z) G(z) \delta 
\rho/\rho$. The use of a log-normal density field has the added benefit of accounting for a small amount of non-linear 
evolution in $\delta \rho/\rho$ and also prevents the unphysical situation $\rho < 0$ (see Coles \& Jones 1991). The choice of the 
5$h^{-1}$\Mpc~cell size reflects a compromise between ensuring that it is: (i) large enough to keep $\delta \rho/\rho$ in 
the linear or quasi-linear regime; (ii) large enough to keep the total number of cells in the simulation volume within computer 
memory constraints; (iii) small enough to have sufficient mass resolution to identify a cluster of galaxies of mass 
$10^{14} h^{-1}$\Msun~(see section 2.8). The computation of the bias $b(z)$ is addressed in section 2.7.

\subsection{The input cosmology}
We use a spatially flat cosmology with parameters $H_{\rm{0}} = 70$\kmpspMpc, $\Omega_{\rm{M}}=0.3$, $\Omega_{\rm{\Lambda}}=0.7$, $f_{\rm{baryon}}=0.16$, $\sigma_{\rm{8}}=0.74$ and $n_{\rm{scalar}}=1$. The CAMB software 
(Code for Anisotropies in the Microwave Background; Lewis et al.~2000) is used to compute a linear isotropic power spectrum 
complete with BAO and an appropriate transfer function. The resulting $z=0$ power spectrum is sampled to generate Fourier modes
of the density field within the simulation volume, which is then fast Fourier-transformed to yield the $z=0$ dark matter density 
field, $\delta \rho/\rho$, defined on the grid of $5 h^{-1}$\Mpc~cells. The latter is the starting point for the main simulation (section 2.2) and the cluster finding algorithm (section 2.8).

\subsection{Luminosity functions} 

Here we describe the luminosity functions used in the simulation. For all populations, it has 
been unavoidably necessary to extrapolate these in luminosity and redshift beyond the regimes in which they were observationally determined. We therefore expect that for many applications users will need to post-process the source catalogues to implement 
more realistic forms of high-redshift evolution, mostly by applying some form of negative space-density and/or luminosity evolution. Alongside the description of each luminosity function, we indicate what we consider to be the most reasonable form of 
high-redshift post-processing for each luminosity function, based on the observational data available at the present time. 
However, in order not to restrict future applications of the simulations as constraints at high redshift improve in the years 
ahead, the online database tool will give the user the choice between applying either this default 
post-processing option or a user-defined form with space-density evolution and/or luminosity evolution decreasing in power-law or negative exponential form (or no post-processing at all). 

In all cases, the radio emission has been modelled without including polarization properties and without making any explicit 
allowance for the effects of the increasing energy density of the cosmic microwave background radiation (CMB) at high redshift. We have also made no attempt to model the Sunyaev-Zeldovich effect on either cluster or galactic-scales. \\

\noindent
$\bullet$ {\em Radio-quiet AGN.}
Classical radio-loud AGN constitute only a small fraction of the overall AGN population; a more complete 
global census of AGN activity is provided by the hard X-ray AGN luminosity function (HXLF). Using the 
latter in conjunction with a radio:X-ray luminosity relation, Jarvis \& Rawlings~(2004) showed that 
radio-quiet AGN can make a significant contribution to the well-characterised upturn in the 1.4~GHz 
source counts below 1~mJy, a feature which had previously been ascribed largely 
(although not without controversy) to star-forming galaxies (see e.g. Seymour et al.~2004). Recent 
multi-wavelength identification work on several deep fields appears to substantiate this claim (e.g. 
Simpson et al.~2006; Seymour et al.~2008; Smolcic et al.~2008), suggesting that possibly as many as half
the sources at the level of a few tens of \microJy~may be radio-quiet AGN. 

We therefore follow the Jarvis \& Rawlings~(2004) prescription for including radio-quiet AGN in our 
simulation. We use the Ueda et al.~(2003) AGN HXLF for the instrinsic (i.e. absorption-corrected) 
2--10\keV~band and their model for luminosity-dependent density evolution (LDDE). Since the Ueda et al.
HXLF does not include Compton-thick AGN (i.e. those with obscuring column densities 
$N_{\rm{H}} > 1.5 \times 10^{24}$\psqcm), we increase the space density given by this luminosity function 
by 50~per cent (independent of X-ray luminosity) in a notional attempt to account for such sources. 
Although there is now compelling evidence from observations (e.g. Risaliti et al.~1999 at low redshift; 
Martinez-Sansigre et al.~2007 at high redshift) and X-ray background models (e.g. Gilli et al.~2007) that 
obscured Compton-thick AGN may be as abundant as unobscured AGN, the precise breakdown as a function of 
intrinsic luminosity and redshift is not well-constrained. We therefore do not believe it worthwhile to 
attempt a more sophisticated correction for Compton-thick AGN at this stage and instead leave the 
matter open for refinement during post-processing.

The intrinsic X-ray (2--10\keV;\ergps) and radio luminosities (1.4~GHz; W/Hz/sr) are assumed to be related 
as follows, as inferred from the correlation found by Brinkmann et al.~(2000) for radio-quiet quasars:

\begin{equation}
\log(L_{\rm{2-10 keV}}) = 1.012\log(L_{\rm{1.4 GHz}})  + 21.3
\end{equation}

\noindent
The correlation of Brinkmann et al. exhibits scatter of $\sim \pm 1$~dex at a given X-ray 
luminosity (and is further biased by the fact that it refers to a sample detected in both soft X-rays 
and at 1.4~GHz), but the extent to which this scatter is intrinsic to the physics of the emission process 
is not clear. It may be largely due to observational errors, coupled with uncertain corrections for X-ray 
absorption and the effects of Doppler de-boosting of radio-core emission. However, the fact that the 
slope of eqn.~(2) is so close to unity strongly argues that this radio emission is a direct tracer of the
nuclear activity, and we therefore follow Jarvis \& Rawlings~(2004) in assuming a 1:1 relation with no 
scatter.

The adopted luminosity limits of log(L$_{\rm{1.4~GHz}}$)=18.7 and 25.7 [W/Hz/sr] span the range from low-luminosity AGN to the most 
luminous quasars. The sources are modelled as point sources with power-law radio spectra of the form $F_{\rm{\nu}} \propto \nu^{-0.7}$ (see e.g. Kukula et al.~1998). The Ueda et al. dataset constrains the HXLF only out to $z \sim 3$ and assumes that the space density of all sources declines as $(1+z)^{-1.5}$ beyond $z=1.9$, and this is assumed for the simulation. However, recent constraints from the Chandra Mult-wavelength project (CHAMP) (Silverman et al.~2007) have pushed the high-redshift constraints out to $z \sim 5$ and suggest that the decline in space density above $z=1.9$ is steeper: $(1+z)^{-3.27}$. Our default post-processing option is therefore to reduce the space density by a factor $(1+z)^{-3.27}$ above this redshift.\\

\noindent
$\bullet$ {\em Radio-loud AGN.}
For the classical double-lobed radio-loud AGN, we use the 151~MHz luminosity function derived by 
Willott et al.~(2001) from a compilation of low-frequency-selected samples. In particular, we use their
`model C' luminosity function (adapted to our chosen cosmology) which consists of low and high luminosity
components with different functional forms and redshift evolutions. The low-luminosity component consists of 
a Schechter function with an exponential cut-off above log L$_{\rm{151~MHz}} \simeq 26.1$~[W/Hz/sr] 
(in the $\Omega_{\rm{M}}=1$ cosmology of Willott et al.) and space density evolution of the form $(1+z)^{4.3}$ out to 
$z=0.706$ and constant thereafter. Willott et al. were not able to place constraints on any high-redshift decline in this
low-luminosity component of the luminosity function and our simulation therefore does not incorporate one. As a default 
post-processing option, however, we suggest a decline in the space density as $(1+z)^{-2.5}$ above $z=2.5$, matching that 
inferred by Jarvis \& Rawlings~(2000) for flat-spectrum quasars.

The high luminosity component consists of 
Schechter function with an inverted exponential cut-off at low luminosity, giving rise to a function which peaks at 
log L$_{\rm{151~MHz}} \simeq 27$~[W/Hz/sr] (in the Willott et al. cosmology). The redshift evolution is modelled as a Gaussian 
peaking at $z = 1.91$ and no further post-processing is proposed or deemed necessary. These properties of the luminosity function 
make it natural to identify the low and high luminosity components crudely with FRI and FRII radio sources, respectively. 
We note, however, that Willott et al. pointed out that the luminosity at which the low- and high-luminosity components 
contribute equally to the luminosity function tends to be about 1~dex higher than the traditional FRI/FRII break luminosity. 
The adopted limits of integration for the luminosity function are: log L$_{\rm{151~MHz}} = 20$ and 28~[W/Hz/sr] for the FRIs, and log L$_{\rm{151~MHz}} = 25.5$ and 30.5~[W/Hz/sr] for the FRIIs 
(all in the simulation cosmology).
\\

\noindent
$\bullet$ {\em Star-forming galaxies.}
For star-forming galaxies, we essentially use the 1.4~GHz luminosity function derived from the IRAS 2Jy sample by Yun, 
Reddy \& Condon~(2001). This is modelled as a sum of two Schechter functions, with the high luminosity component dominant 
at log L$_{\rm{1.4~GHz}} > 22 $~[W/Hz/sr], which equates to a 60\micron~luminosity of approximately $10^{11}$\Lsun. For the 
purposes of assigning morphologies and SEDs, we follow Yun, Reddy \& Condon and identify these low and high luminosity components 
with normal/quiescent late-type galaxies and starburst galaxies, respectively. This separation is borne out by the analysis of 
Takeuchi et al.~(2003), who remeasured the IRAS 60\micron~luminosity function and showed that these low and high luminosity 
components correspond to the cool and warm subpopulations (defined to have $\beta=S(100\micron)/S(60 \micron)$~flux ratios 
$\beta > 2.1$ and $< 2.1$) respectively. 

We employ lower and upper luminosity limits of log L$_{\rm{1.4~GHz}} =17 $ and 
25.5~[W/Hz/sr] for the normal galaxies, and log L$_{\rm{1.4~GHz}} = 21 $ and 27~[W/Hz/sr] for the starbursts. 
The behaviour of the faint end of the radio luminosity function for the star-forming galaxies is a controversial but critical 
issue. The observational data of Yun, Reddy \& Condon~(2001) clearly diverge below their Schechter function fit below 
log L$_{\rm{1.4~GHz}} =19.3$~[W/Hz/sr], an effect the authors ascribe to measurement incompleteness. The star-forming galaxy 
luminosity function of Sadler et al.~(2002), based on cross-matching the 1.4~GHz NVSS with the 2dFGRS, also peaks around 
log L$_{\rm{1.4~GHz}} = 19$~[W/Hz/sr]~and may decline towards lower luminosities. However, other determinations of the
luminosity function with higher degrees of faint end completeness (e.g. Mauch \& Sadler~2007, Condon et al.~2002) show instead
that the luminosity function just flattens below log L$_{\rm{1.4~GHz}} \sim 20$~[W/Hz/sr], with no evidence for an actual downturn.
To take account of this, we simply assume that the $z=0$ Yun, Reddy \& Condon luminosity function is flat below 
log L$_{\rm{1.4~GHz}} = 19.6$~[W/Hz/sr]. It has been shown that the local 1.4~GHz~luminosity function of normal galaxies 
is (when suitably transformed) indistinguishable from the luminosity function of 60\micron-selected galaxies 
(e.g. Condon~1989,1992). We therefore assume that the radio emission serves as a reliable tracer of all star formation 
(albeit subject to the uncertainty in the faint end of the radio luminosity function), and do not follow 
the example of Hopkins et al.~(2000)/Windhorst~(2003), who supplemented the warm IRAS luminosity function in their sky simulations with 
an additional normal galaxy population derived from an optical luminosity function.

For both components of the luminosity function we assume a form of cosmological evolution which, in a Universe with 
$\Omega_{\rm{M}}=1$ and $\Omega_{\rm{\Lambda}}=0$ takes the form of pure luminosity evolution of the form $(1+z)^{3.1}$ 
out to $z=1.5$, with no further evolution thereafter. This is adapted to the cosmology of interest by shifting the 
luminosities and scaling the space density inversely with the ratio of the differential co-moving volumes, dV/dz, for the two cosmologies. Evolution of essentially this form (deduced for an $\Omega_{\rm{M}}=1$ Universe) was inferred from studies of IRAS-selected galaxies (Rowan-Robinson et al.~1993) and for sub-mm-selected galaxies (Blain et al.~1999). More recent determinations of the
evolution of the radio luminosity function of star-forming galaxies are consistent with this, mostly finding luminosity 
evolution of the form $(1+z)^{Q}$ ($Q=2-3$) and negligible density evolution (e.g. Seymour et al.~2004; Hopkins~2004). 
{\em Spitzer} results on the evolution of the infrared (8--1000\micron) and far-infrared luminosity functions reveal a 
degeneracy between luminosity and density evolution which is consistent with our adopted form of evolution (Le Floc'h et 
al.~2005; Huynh et al.~2005), although constraints on the population with $L_{\rm{IR}} < 10^{11}$\Lsun~(equivalent to 
log $L_{\rm{1.4 GHz}}=22$ [W/Hz/sr]) are limited at $z > 0.5$. 

More complex models of the evolution of star-forming galaxies have been put forward to interpret mid- and far-infrared 
observations (e.g. Franceshini et al.~2001; Lagache et al.~2004). Many such models break the population down into 
contributions from weakly or non-evolving normal galaxies, together with a population of starburst galaxies subject to strong 
evolution in space density and luminosity. The model of Franceshini et al.~(2001) is one such model, which takes as its 
starting point a 12\micron~local luminosity function. However, when translated to 60\micron~with the assumed infrared SEDs, 
the resulting breakdown into normal and starburst galaxies (Fig.~11 of Franceschini et al.~2001) is completely different from 
the one we assume, with the starbursts comprising a decreasing fraction of the whole for $L_{\rm{60}} > 10^{10}$\Lsun. 
When translated to 24\micron~(Gruppioni et al.~2005), the Franceschini et al.~model is at variance with the 
redshift breakdown of {\em Spitzer} source counts reported by Le Floc'h et al.~(2005). The Lagache et al.~(2004) model decomposition of the local 60\micron~luminosity function into normal and starburst galaxies is similar to ours, but again there are 
discrepancies with respect to the {\em Spitzer} luminosity functions reported by Le Floc'h et al.~(2005). In the absence of
a definitive alternative model and to avoid the uncertainties arising from K-corrections in the mid-infrared, we prefer to base 
our predictions starting from the local 1.4~GHz luminosity function plus pure luminosity evolution for both sub-populations, as 
already discussed . 

The simulation was performed by assuming no further evolution in the luminosity function beyond redshift $z=1.5$. 
Taken at face value this is clearly unrealistic, but it gives the user freedom to impose their own form of high redshift 
decline by selectively filtering the catalogue during post-processing. At the present time, observations show that 
the star formation rate density is essentially constant from $z \sim 1.5$ to $z \sim 4$ and then to drops off sharply above 
this redshift (see e.g. Hopkins \& Beacom~2006). As our default post-processing option, we model this high-redshift fall-off 
using the piece-wise power-law model of Hopkins \& Beacom, in which the star-formation rate density falls off as $(1+z)^{-7.9}$ 
above $z=4.8$.\\

\noindent
$\bullet$ {\em AGN/star-forming hybrid galaxies and double counting.}
The interplay between bursts of star formation and black hole accretion is of considerable interest for studies of galaxy 
evolution, and one aspect of this concerns hybrid galaxies in which both processes make significant contributions to the observed 
radio emission. Unfortunately, such objects cannot be isolated in the simulations due to the use of separate luminosity functions 
for the AGN and the starbursts, but they are almost certainly accounted for by virtue of the way in which the luminosity functions 
were constructed. A fraction of the luminous starbursts are likely to be mis-classified obscured AGN (or at least have a 
significant AGN contribution), and similarly some of the radio emission in the low luminosity AGN (especially the radio-quiet 
population) is likely to be related to star formation. The most likely upshot of this is that, in the simulation as a whole, the
latter populations may have been double counted to a certain extent, thereby slightly overproducing the source counts. The 
semi-analytical simulations, in contrast, will include hybrid galaxies. A further instance of potential double counting 
concerns the fact that the hard X-ray luminosity function used for the radio-quiet AGN implicitly includes some contribution 
from the radio-loud AGN, so that the latter are in effect counted twice. However, this effect is likely dwarfed by the 
uncertainty in the Compton-thick correction factor applied to the X-ray luminosity function and we therefore make no further allowance for it. \\

\subsection{Radio-loud AGN: Unification, beaming, morphologies and radio spectra}
As described in section 2.4, radio-loud AGN are initially drawn from a 151~MHz luminosity function for steep-spectrum, lobe-dominated 
sources. We use this as the input parent population for an orientation-based unification model which enables us to assign source 
structures and radio spectra in a physically-motivated manner. Some of this procedure follows the prescription of Jackson \& 
Wall~(1999). The steps in our process are as follows: \\

\noindent
(i) Sources are assigned
a true linear size, $D_{\rm{true}}$, drawn at random from a uniform distribution [0,$D_{\rm{0}} (1+z)^{-1.4}$], 
where $D_{\rm{0}}=1$~Mpc; this assumes that the sources expand with uniform velocity until they reach a size equal to the 
redshift-dependent upper envelope of the projected linear size distribution measured by Blundell, Rawlings \& Willott~(1999). \\

\noindent
(ii) The angle between the jet axis and the observer's line-of-sight is 
drawn from a uniform distribution in cos $\theta$, and the jet axis is given a random position angle on the sky. \\

\noindent
(iii) The ratio of the intrinsic core to extended luminosities, defined at
1.4~GHz in the rest-frame, is given by $R_{\rm{CL}}=10^x$, where $x$ is drawn from a Gaussian distribution of mean $x_{\rm{med}}$ 
and $\sigma=0.5$, truncated at abs$(x)>10$ to avoid numerical problems. The numerical values of $x_{\rm{med}}$ we use are given at the end of this subsection.\\

\noindent
(iv) A relativistic beaming model is used to derive the 
observed core:extended flux ratio, $R_{\rm{OBS}}= R_{\rm{CL}} B(\theta)$, where $B(\theta) = \frac{1}{2}[(1 - \beta cos \theta)^{-2} + 
(1 + \beta cos \theta)^{-2}]$; $\beta= \sqrt(\gamma^2 - 1)/\gamma$ and $\gamma$ is the Lorentz factor of the jet. \\

\noindent
(v) The extended
emission is modelled with a power-law spectrum $F_{\rm{\nu}} \propto \nu^{-0.75}$, while the core spectrum is modelled with some 
curvature: $\log F_{\rm{\nu}} =  a_{\rm{c0}} + a_{\rm{c1}}\log \nu + a_{\rm{c2}} (\log \nu)^{2}$. For $\nu$ in GHz, $a_{\rm{c1}}=0.07$,
$a_{\rm{c2}}=-0.29$, as measured by Jarvis \& Rawlings~(2000) from a sample of flat spectrum quasars ($a_{\rm{c0}}$ sets the normalization). The observational data on which these fits were based typically extend up to 10 or 20 GHz, implying that the model SEDs cannot be simply extrapolated to higher frequencies (e.g. to the WMAP bands above 20 GHz). \\

\noindent
(vi) For FRIs, the extended flux distribution on the sky is modelled as two coaxial elliptical lobes of uniform surface brightness, 
extending from the point source core, and each with a major axis length equal to half the projected linear size. The axial ratio
of the lobes is drawn from a uniform distribution [0.2,1]. \\

\noindent
(vii) For FRIIs, the inner edges of the lobes are offset from the core by a distance $f\times PLS$, where
$PLS$ is the projected linear size of the whole source and $f$ is drawn from the uniform distribution [0.2,0.8]. A fraction $f_{\rm{HS}}$ of the extended flux in the FRIIs is assigned to point-source hotspots positioned at the ends of the lobes, where 

\begin{equation}
f_{\rm{HS}}= 0.4[\log L_{\rm{151~MHz}} - 25.5] \pm 0.15,
\end{equation}
in accordance with a correlation found by Jenkins \& McEllin~(1977) (the scatter is modelled with a uniform distribution). The 
hotspots are assumed to have the same radio spectral shape as the rest of the lobe.  \\

\noindent
(viii) In an attempt to model self-absorption within compact sources (the so-called GigaHertz Peaked Spectrum (GPS) sources) we apply 
a spectral turnover at a frequency, $\nu_{\rm{p}}$, below which the spectrum asymptotes to $F_{\rm{\nu}} \propto \nu^{2.5}$. We use 
the following relation, found by O'Dea~(1998), between $\nu_{\rm{p}}$ and $D_{\rm{true}}$: 

\begin{equation}
\log \nu_{\rm{p}} (\rm{GHz}) = -0.21  - 0.65\log D_{\rm{true}} (\rm{kpc}). \\
\end{equation} 

The above formalism should be able to reproduce the observed variety of radio-loud AGN, including radio 
galaxies, steep and flat-spectrum spectrum radio quasars, and strongly beamed sources such as BL Lacs and blazars. 
The key parameters of the model are the core:lobe ratios ($x_{\rm{med}}$) and jet Lorentz factors ($\gamma$) for the FRI
and FRII parent populations, and their possible dependence on $L_{\rm{151~MHz}}$. There is a wealth of literature which 
could in principle be used to constrain these parameters, but none of it is ideally suited to our needs due to a combination 
of selection effects and poorly defined asumptions. Instead, we have tuned the parameters to reproduce the observed multi-frequency
radio source counts (as shown in section 3.2) at flux densities $>$ 1 mJy where such sources dominate. 
We find $(x_{\rm{med}}, \gamma) = (-2.6,6)$ and $(-2.8,8)$ for the FRI and FRIIs, respectively. For simplicity, we do not incorporate any luminosity dependence in these parameters. 

\subsection{Star-forming galaxies: source sizes, radio spectra and HI}
As described in section 2.4, we split the star-forming galaxies into two sub-populations of `normal' galaxies and `starburst' 
galaxies, according to whether they are drawn from the low or high luminosity Schechter function component of the Yun, Reddy \&
Condon~(2001) luminosity function, respectively. Here we discuss the assignment of radio spectra and structural properties 
to these populations.

As reviewed by Condon~(1992), the radio spectra of star-forming galaxies consist of two components: (i) thermal free-free emission
with a luminosity directly proportional to the current photoionization rate and $F_{\rm{\nu}} \propto \nu^{-\alpha}$ with $\alpha \sim 0.1$; (ii) non-thermal synchrotron emission from supernovae for which the power-law slope $\alpha_{\rm{nt}} = 0.75 \pm 0.1$. The ratio of the total luminosity to that of the thermal component is not well constrained but is commonly assumed to be $\sim 1 + 10(\nu_{\rm{GHz}})^{0.1- \alpha_{\rm{nt}}}$. In the presence of free-free absorption, acting on both the thermal and non-thermal components, the radiative-transfer solution for the overall radio spectrum takes the form:  
 
\begin{equation}
L_{\rm{\nu}} \propto \nu^{2} [1 - e^{-\tau_{\rm{ff}}}][1 + 10\nu_{\rm{GHz}}^{0.1-\alpha_{\rm{nt}}}].
\end{equation}

The free-free opacity, $\tau_{\rm{ff}}$, scales with frequency as $\tau_{\rm{ff}} = (\nu_{\rm{ff}}/\nu)^{2.1}$ where $\nu_{\rm{ff}}$ is
a normalisation parameter. For most normal galaxies $\nu_{\rm{ff}} = 0.003-0.01$~GHz and for the observed frequencies of interest 
eqn.~(5) can be recast as the sum of two power-laws:

\begin{equation}
L_{\rm{\nu}} \propto \nu_{\rm{GHz}}^{-0.1} + 10\nu_{\rm{GHz}}^{-\alpha_{\rm{nt}}}.
\end{equation}

For starburst galaxies, the more compact sizes and higher gas densities result in higher $\nu_{\rm{ff}}$ values and we use 
$\nu_{\rm{ff}}=1$~GHz, as measured in the central regions of M82. For completeness, we add a far-infrared thermal dust component
to the spectral energy distribution of star-forming galaxies using a modified black-body with $T=45$\K~and $\beta=1.5$. The dust 
SED is normalised using the far-infrared:radio correlation to relate the $60$\micron~and 1.4~GHz~luminosities (see Yun, Reddy \&
Condon~2001). This component has a discernable impact on the observed SED only at very high redshift $z > 15$ and at the highest output frequency (18~GHz).

The assignment of disk sizes to the normal galaxies invokes a chain of reasoning which begins with the remarkably tight relation between HI mass ($M_{\rm{HI}}$) and disk size ($D_{\rm{HI}}$, in \kpc, defined at an HI density of 1\Msun~pc$^{-2}$) measured by Broeils \& Rhee~(1997) for a 
sample of local spiral and irregular galaxies:

\begin{equation}
\log M_{\rm{HI}} = (1.96 \pm 0.04)\log D_{\rm{HI}} + (6.52 \pm 0.06).
\end{equation}

We relate $M_{\rm{HI}}$ to $L_{\rm{1.4 GHz}}$ by combining the star formation rate-$L_{\rm{1.4 GHz}}$ relation of Sullivan et al.~(2001) with the looser correlation between star formation rate and M$_{\rm{HI}}$ (Fig.~3, left panel in Doyle \& Drinkwater~2006) to yield:

\begin{equation}
\log M_{\rm{HI}} = 0.44\log L_{\rm{1.4 GHz}}\rm{[W/Hz/sr]} + 0.48 \pm \Delta, 
\end{equation}
where $\Delta$ is random scatter drawn from a normal distribution with $\sigma=0.3$. We derive a fiducial radio continuum disk diameter, $D_{\rm{cont}}$, using the relationship given by Broeils \& Rhee~(1997) between $D_{\rm{HI}}$ and the optical absorption-corrected diameter, $D_{\rm{25}}^{\rm{b,i}}$ (measured at the 25th mag~arcsec$^{-2}$ isophote): 
 
\begin{equation}
\log D_{\rm{cont}} = \log D_{\rm{HI}} - 0.23 - \log(1+z).
\end{equation}

The $(1+z)$ factor is not from Broeils \& Rhee~(1997) but an approximate scaling of disk diameter with redshift suggested by cosmological
hydrodynamic simulations of disk formation (C.~Power, private communication). For comparison, we note that Ferguson et al.~(2004) suggest that galaxy radii scale as the inverse of the Hubble parameter, $H(z)^{-1}$, as deduced from rest-frame $UV$ observations with
the {\em Hubble Space Telescope}.

For the starbursts, we assume the following continuum sizes:

\begin{equation}
D_{\rm{cont}} = (1+z)^{2.5} {\rm kpc},
\end{equation}
out to $z=1.5$ and constant at 10~kpc for $z>1.5$. This reflects the fact that local starbursts are compact (kpc-scale) whereas $z \sim 2$ sub-mm galaxies are an order of magnitude larger in scale (see e.g. MERLIN interferometry, Chapman et al.~2004). Equation (8) is again used to assign a nominal HI mass to the starbursts, although the relation is unlikely to hold for such systems. In the output catalogues, star-forming disks are placed on the sky with random orientations. It should be noted that galaxies are catalogued in the output only according to their spatially integrated fluxes, and not on the basis of surface brightness. Users may thus model the source sizes and intensity profiles in a more sophisticated way during post-processing, should they so wish.

\subsection{Large-scale clustering and biasing}

The bias $b(z)$ which appears in eqn.~(1) is computed separately for each galaxy population using the formalism of 
Mo \& White~(1996). We make the simple assumption of assigning each population an effective dark matter halo mass 
which reflects its large-scale clustering. Our simulation does not have the mass resolution to directly 
resolve galaxy and group-sized haloes, so we are not literally assigning galaxies to individual haloes; our approach is thus distinct from the technique based on the `halo occupation distribution function' which is 
used to populate dark matter haloes in N-body simulations, e.g. Benson et al. (2000). Whilst the assumption of a fixed halo mass for an entire population is reasonable at low redshifts, the formalism breaks down towards the higher redshifts of the simulation
as $b(z)$ increases, leading to a potential blow-up in the exponent of eqn.~(1). To circumvent the excessively strong 
clustering which would otherwise result, the bias for each population is held constant beyond a certain cut-off redshift ($z_{\rm{cut}}$). Halo 
masses are assigned to the populations as follows: \\

\noindent
$\bullet$ {\em Radio-quiet AGN.} 
The clustering of quasars and its redshift evolution have been measured by the 2dFQZ survey (e.g. Croom et al.~2005) and
found to be well described by a constant halo mass of $3 \times 10^{12} h^{-1}$\Msun~out to redshift limits of the 2dFQZ ($z=2.5$). 
We thus adopt this mass and set $z_{\rm{cut}}=3$. \\

\noindent
$\bullet$ {\em Radio-loud AGN.} 
The most extensive information on the cosmological clustering of radio sources has been derived from measurements of
the angular clustering in the {\em NVSS} and {\em FIRST} radio surveys (see e.g. Overzier et al.~2003 and references
therein). With assumptions about the redshift distribution for the radio sources, these clustering measurements can be
inverted to yield 3-D correlation lengths and dark halo masses. We use the discussion in Overzier et al.~(2003) to fix 
our halo masses. For FRIs, we take $M_{\rm{halo}}=10^{13} h^{-1}$\Msun, which reflects the clustering of low-luminosity radio 
sources and $L \sim L^{\star}$ E-type galaxies. For FRIIs, we take $M_{\rm{halo}}=10^{14} h^{-1}$\Msun, which reproduces the 
clustering of the powerful radio sources and matches the clustering evolution of z=0 $L \gtrsim L^{\star}$ E-type galaxies, $z \sim 1$ EROs and 
$z \sim 0.55$ LRGs. These masses are comparable to a small group and a small
cluster, respectively, and are consistent with the findings of other studies which show that radio sources tend to reside in 
overdense, highly-biased environments. Since the resulting $b(z)$ increases very rapidly with redshift, which would lead to
an excessively inhomogenous clustering if left unchecked, we impose $z_{\rm{cut}}=1.5$. This is also the effective cut-off in
the redshift distribution of clusters of galaxies (see next subsection), and the redshift beyond which, in any case, existing radio surveys do 
not constrain the clustering.  \\

\noindent
$\bullet$ {\em Star-forming galaxies.} For the normal galaxies, we take $M_{\rm{halo}}=10^{11} h^{-1}$\Msun~and $z_{\rm{cut}}=3$ 
which, as discussed by Overzier et al.~(2003), reproduces the z=0 clustering of IRAS galaxies, L $\sim$ 0.5 L* late-type galaxies and LBGs at $z \sim 3$. This population has a bias close to unity. For the starbursts, we take $M_{\rm{halo}}=5 \times 10^{13} h^{-1}$\Msun, consistent with the claim of 
Swinbank et al.~(2006) that submillimetre galaxies are progenitors of $L > 3L^{\star}_{\rm{K}}$ E-type galaxies at z=0, and with the clustering measurements of Farrah et al.~(2006) for mid-IR selected samples at $z=1.5-3$. For such a group/cluster-sized mass, 
we again impose $z_{\rm{cut}}=1.5$. We note, however, that tuning the starburst clustering to match the high-redshift submillimetre population will not accurately reproduce the much weaker clustering of the local starburst galaxies, which is much lower than that of passive galaxies (e.g. Madgwick et al.~2003). \\
 
\subsection{Clusters of galaxies}
The simulation was designed for modelling large-scale structure on $\gg 5 h^{-1}$\Mpc~scales. As discussed in section 2.7, 
the radio-loud AGN are heavily biased with respect to the underlying dark matter density field and are expected to lie preferentially in group/cluster environments. It was thus considered advantageous to be able to identify the clusters of galaxies in the simulation 
directly, since they are an important population in their own right for investigations of cosmic magnetism and the 
Sunyeav-Zeldovich effect, for example. We attempted to do this whilst ensuring consistency with the framework for modelling 
large-scale structure outlined in section 2.2.

This goal was accomplished through the use of a numerical Press-Schechter (Press \& Schechter~1974) method to identify the cluster-sized haloes. The basic 
assumption is that any given cell is considered to be part of the largest mass halo which could have collapsed by the epoch under consideration. Under linear theory, a halo is defined to be collapsed when its extrapolated linear overdensity reaches the value 
$\delta \rho /\rho = \delta_{\rm{crit}} = 1.66$. \footnote{In the standard spherical collapse Press-Schechter model, $\delta_{\rm{crit}}$ 
depends only weakly on the cosmological parameters -- see e.g. Eke et al.~(1996) -- and is assumed to be independent of scale; Sheth et al.~(2001) describe how an 
ellipsoidal collapse model with a scale- (i.e. mass) dependent threshold can improve the agreement with the mass function derived
from N-body simulations, and that approximately the same results can be obtained in the spherical collapse model by reducing $\delta_{\rm{crit}}$ by $\simeq 16$~per cent.} We thus filter the dark matter density field, $\delta \rho /\rho$, on a range 
of successively smaller scales and at each filter step flag all pixels for which the smoothed density field exceeds 
$\delta_{\rm{crit}}/G(z)$ (where $G(z)$ is as before the linear growth factor for the cell of interest, since we are once again 
looking out into the light cone). The next step is to locate the islands of inter-connected overdense cells and identify them 
as collapsed haloes, with a mass equal to that contained within the filter volume for the mean density of the Universe. The 
process is then repeated at the next lower mass scale, with the proviso that any cell already part of a larger mass structure 
is ignored. 

\begin{figure}
\includegraphics[width=0.48\textwidth,angle=0]{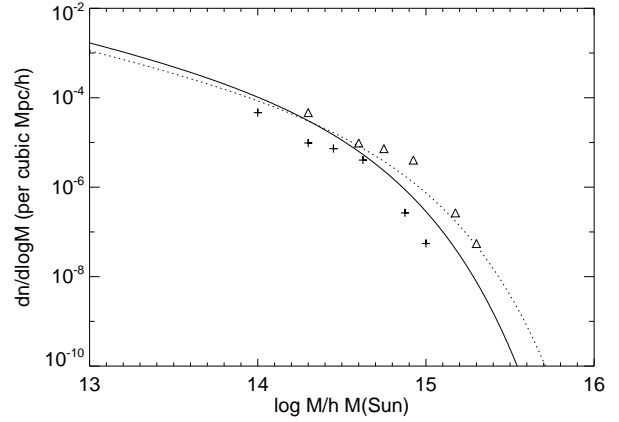}
\caption{\normalsize The crosses show the simulated cluster mass function derived from the `quantised' cluster masses in a $10^{9}h^{-3}$\Mpc$^{3}$ volume at $z=0$. The solid and dotted lines show the Press-Schechter~(1974) and Sheth-Tormen~(1999) mass functions, 
respectively. It is clear that some additional boosting of the masses is necessary in post-processing when converting the quantised
masses into a continuous distribution. The open triangles show the effect of artificially boosting all masses by a factor of 2, 
as discussed in section 2.8.}
\label{fig:clustermf}
\end{figure}

The finite cell size of the simulation introduces a discreteness in the filter size steps and hence in the masses of the resulting
clusters, although this becomes fractionally smaller at higher masses. The smallest mass filter includes seven 
$5 h^{-1}$\Mpc~cells (corresponding to a cluster mass of $7.26 \times 10^{13} h^{-1}$\Msun) and thereafter the filter grows 
in a quasi-spherical manner whilst retaining symmetry along the $xyz$ axes of the simulation volume. The resulting cluster masses 
are quantised to the following values: $[0.726, 1.97, 2.80, 3.42, 5.91, 8.4] \times 10^{14} h^{-1}$\Msun. This quantisation of the 
cluster masses makes it very difficult to compare with continuous analytical cluster halo mass functions and observations of cluster counts; nevertheless, we attempted to do this by running the cluster-finder algorithm on a $10^{9}h^{-3}$\Mpc$^{3}$ volume at $z=0$, 
assuming that the quantised masses are minimum estimates and that the true cluster masses actually lie in a range up to the next highest mass level. The resulting mass function is shown in Fig.~\ref{fig:clustermf} for comparison with Press-Schechter~(1974) and Sheth-Tormen~(1999) mass functions; our mass function generally falls short of the latter and some boosting of the masses will be necessary when post-processing the quantised cluster masses to generate a continuous mass distribution. As a crude example, we show that increasing all masses by a factor of 2 would bring the mass function into better agreement with the Sheth-Tormen function. A full investigation into the origin of this discrepancy is beyond the scope of this paper, and we simply note that it may result from a 
combination of several effects such as: (i) the use of a sharp-edged (top-hat) smoothing function for filtering the density 
field and the neglect of infall regions and inter-cell interpolation at the edge of the filter (e.g. the spherical filter enclosing
a mass of $10^{15}h^{-1}$\Msun~has a radius of $14.2 h^{-1}$\Mpc, corresponding to just 2.8 cells, so systematic filter edge effects 
are likely to be significant); (ii) the fact that the appropriate mass density for the collapsed objects may slightly exceed the assumed 
value (the mean density of the Universe); (iii) the reasons discussed by Sheth et al.~(2001) for favouring an ellipsoidal collapse model -- or, equivalently, a lower spherical collapse threshold -- over the standard spherical Press-Schechter formalism.

\begin{figure}
\includegraphics[width=0.48\textwidth,angle=0]{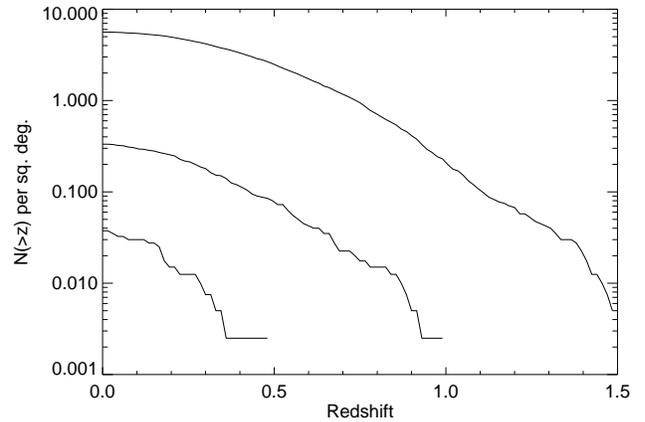}
\caption{\normalsize Surface densities of galaxy clusters lying at redshift $z$ or higher, exceeding various mass thresholds. 
From top to bottom, the three lines correspond to mass thresholds of $7.0 \times 10^{13}$, $2.0 \times 10^{14}$, and $3.4 \times 10^{14} h^{-1}$\Msun, respectively. When the `quantised' masses are artificially boosted by a factor of 2 as described in Fig.~\ref{fig:clustermf}, the same three lines now nominally correspond to mass limits $1.4 \times 10^{14}$, $4.0\times 10^{14}$ and $6.8 \times 10^{14} h^{-1}$\Msun.}
\label{fig:zclus}
\end{figure}

The areal density of clusters as a function of redshift is shown in Fig.~\ref{fig:zclus}, which can be compared in principle 
with observational results (e.g. van Breukelen et al.~2006) and N-body simulations (Evrard et al.~2002). However, the quantisation 
of the catalogued cluster masses again makes comparisons difficult. A proper comparison will only be possible once 
conversion to a continous mass function has been carried out in post processing. Nevertheless, we note the results of van 
Breukelen et al.~(2006), who report a surface density of $\sim 10$~clusters deg$^{-2}$ for $z=0.5-1.5$ and masses $>10^{14}$\Msun. 
By way of comparison, our total cluster catalogue shows a density of 5.6~clusters deg$^{-2}$ above the `boosted' mass of $1.4 \times 10^{14} h^{-1}$\Msun, 50 per cent of which are at $z>0.5$. 

Having identified the clusters in the dark matter density field, we used the following procedure to populate them with 
galaxies. Any radio-loud or radio-quiet AGN in a cell within the smoothing filter radius of an identified cluster is assumed
to be a member galaxy of that cluster. Star-forming galaxies of both types are placed without regard for the presence 
of clusters, i.e. just as described in section 2.2. We did not attempt to model the phase space distribution of cluster 
galaxies in a detailed manner; rather, galaxies were assigned a random orientation with respect to the cluster centre and the 
physical radius was drawn at random from the uniform distribution $[0,0.5r_{\rm{virial}}]$, where $r_{\rm{virial}}$ is the cluster
virial radius. The latter was calculated from the cluster mass, $M_{\rm{cl,14}}$ (in units of $10^{14} h^{-1}$\Msun), on the common assumption that the mean density within the cluster is $\sim 100$ times the background density of the universe ($=3 H_{\rm{0}}^{2} \Omega_{\rm{M}} (1+z)^{3}/8 \pi G$) (see e.g. Eke et al.~1996):

\begin{equation}
r_{\rm{virial}} = 1.4 M_{\rm{cl,14}}^{1/3} (1+z)^{-1} h^{-1} \rm{Mpc} 
\end{equation}

This is equivalent to assuming a 3-D galaxy density profile of the form $\rho \propto r^{-2}$ within $r_{\rm{virial}}$. 
Galaxies are offset in velocity from the cluster redshift by a random velocity drawn from a Gaussian distribution with 
standard deviation $0.8\sigma_{\rm{virial}}$. The calculation of the latter draws on the virial theorem ($\sigma_{\rm{virial}}^{2} = G M_{\rm{cl}} /r_{\rm{virial}}$) to yield:

\begin{equation}
\sigma_{\rm{virial}} = 557 M_{\rm{cl,14}}^{1/3} (1+z)^{1/2} \rm{km/s}
\end{equation}

Cluster membership is indicated in the output galaxy catalogues with reference to a separate table listing the cluster properties, giving users the freedom to populate the clusters in a more sophisticated manner during post-processing, should they
so wish.

\subsection{Redshift space distortions}
Within the linear-theory framework that we adopt, it is also possible to include the 
effect of peculiar velocities of galaxies on their measured redshifts. Starting from the 
initial density field with Fourier modes $\delta_{\rm{k}}$, the Fourier modes of the velocity 
field are given by: 

\begin{equation}
\delta \bmath{v_k}(z) = -\frac{iH(z)\delta_k(z)}{k G(z)}\left | \frac{\mathrm{d}G(z)}{\mathrm{d}z}\right | \bmath{\hat{k}}
\end{equation}

\noindent
(Peebles~1976, Kaiser~1987, Peacock~1999). Each component of the peculiar velocity vector field is obtained by taking the Fourier transform of the vector modes in eqn.~(13). The redshift distortion is then calculated at every point in the grid by projecting the peculiar velocity onto the line of sight and perturbing the original Hubble-flow redshift accordingly. When downloading source catalogues from the online database, users will have the choice between selecting a pure Hubble-flow redshift, or a redshift which also includes the induced peculiar motions, the latter being the default option. Note that this formalism does not include the effect on the peculiar velocities of non-linear growth at late cosmological epochs, but it does include the linear-theory distortion of measured clustering discussed by Kaiser~(1987).

\section{SIMULATION OUTPUT AND BASIC TESTS}

Here we describe the output of the simulation and present a few consistency checks to demonstrate that it is in satisfactory agreement with existing observational constraints and can thus be extrapolated safely beyond this regime. The tests were performed independently of the main simulation to provide maximum exposure to potential errors, using the raw simulation output (i.e. with no post-processing applied).

\subsection{Simulation output : source catalogues}

The output of the simulation consists of two inter-linked catalogues, the radio source catalogue and the cluster catalogue, which 
can be accessed through a web-interface at http://s-cubed.physics.ox.ac.uk. Their structures are shown in Tables~1 and 2, 
respectively. Each entry (row) of the radio source catalogue corresponds to a single structure (identified by the source index) and lists the properties of that structure. Each galaxy is made up of one or more such structures, 
each having the same galaxy index, but different source indices, e.g. for each FRI radio source there are three structural components 
(point source core and two elliptical lobes); FRIIs have five components (point source core, two elliptical lobes and two 
point-source hotspots). Radio-quiet quasars and star-forming galaxies all consist of a single structual component. A non-zero
cluster index indicates that the galaxy is part of a cluster of galaxies, and the physical properties of the relevant cluster 
are listed in the cluster catalogue (Table 2). Note that the radio source table structure permits the specification of full polarization information (i.e. $IQUV$ flux densities), but only the intensity $I$ is generated by this simulation. Polarization information (i.e. $QUV$ flux densities) will be supplied later by other members of the SKADS consortium.

The total numbers of sources generated by the simulation are shown in Table~3, broken down into the numbers of galaxies and
sub-structures (the two columns differ only for the FRIs and FRIIs). Note that the figures quoted in this table exceed by a
few per cent those within the canonical $20 \times 20$~deg$^{2}$ simulation area, and that users should take care only to use
sources within the latter area. These unavoidable edge effects arise from the discrete cells used in the simulation and diminish
in importance towards the higher redshifts. The total number of galaxy clusters in the output catalogue is 2248, of which 2201 are within the canonical $20 \times 20$~deg$^{2}$ simulation area.

\begin{table*}
\caption{Structure of the radio source catalogue}
\begin{tabular}{|lll|} \hline
Column & Fortran Format & Content description  \\ \hline
1      & I10    &  Source index \\
2      & I6     &  Cluster index \\
3      & I10	&  Galaxy index \\
4      & I4     &  Star-formation type index (0=no SF, 1=normal galaxy, 2=starburst) \\
5      & I4     &  AGN-type index (0=no AGN, 1=RQQ, 2=FRI, 3=FRII, 4=GPS) \\
6      & I4     &  Structure-type (1=core, 2=lobe, 3=hotspot, 4=SF disk) \\
7      & F11.5	&  RA (degrees) \\
8      & F11.5  &  DEC (degrees) \\
9      & F10.3  &  Comoving distance (Mpc) \\
10     & F10.6  &  Redshift \\
11     & F9.3   &  Position angle (rad) for elliptical substructures \\
12     & F9.3	&  Major axis (arcsec)   \\
13     & F9.3	&  Minor axis (arcsec)   \\
14-17  & F9.4   &  log(I), log(Q), log(U), log(V) flux densities (Jy) @ 151 MHz \\
18-21  & F9.4   &  log(I), log(Q), log(U), log(V) flux densities (Jy) @ 610 MHz \\
22-25  & F9.4   &  log(I), log(Q), log(U), log(V) flux densities (Jy) @ 1400 MHz \\
26-29  & F9.4   &  log(I), log(Q), log(U), log(V) flux densities (Jy) @ 4860 MHz \\
30-33  & F9.4   &  log(I), log(Q), log(U), log(V) flux densities (Jy) @ 18000 MHz \\
34     & F9.4   &  log (M$_{\rm{HI}}$/\Msun) (for star-forming galaxies only) \\
35     & F9.4   &  cos (viewing angle) (relative to jet axis; FRI and FRII only) \\ \hline
\end{tabular} \\ 
\end{table*}

\begin{table*}
\caption{Structure of the cluster catalogue}
\begin{tabular}{|lll|} \hline
Column & Format & Content description  \\ \hline
1      & Long integer & Cluster index  \\
2      & Float        & RA (degrees) \\
3      & Float        & DEC (degrees) \\
4      & Float        & Redshift \\
5      & Float        & Cluster mass ($h^{-1}$ \Msun) \\
6      & Float        & Cluster virial radius ($h^{-1}$ \Mpc) \\
7      & Float        & Cluster velocity dispersion (\kmps)  \\ \hline
\end{tabular} \\ 
\end{table*}

\begin{table*}
\caption{Numbers of galaxies and their structural sub-component sources in the output catalogues of the $20 \times 20$~deg$^{2}$ semi-empirical simulations }
\begin{tabular}{|lll|} \hline
Source type & Number of galaxies ($10^{6}$) & Structural components ($10^{6}$) \\ \hline
Radio-quiet AGN  & 36.1  & 36.1 \\
FRI      &   23.8    & 71.4 \\
FRII    &  0.00235   & 0.012   \\
Normal galaxies     & 207.8   & 207.8  \\
Starburst galaxies      & 7.26     &  7.26 \\ \hline
\end{tabular} \\ 
\end{table*}

\subsection{Multi-frequency source counts}

\begin{figure*}
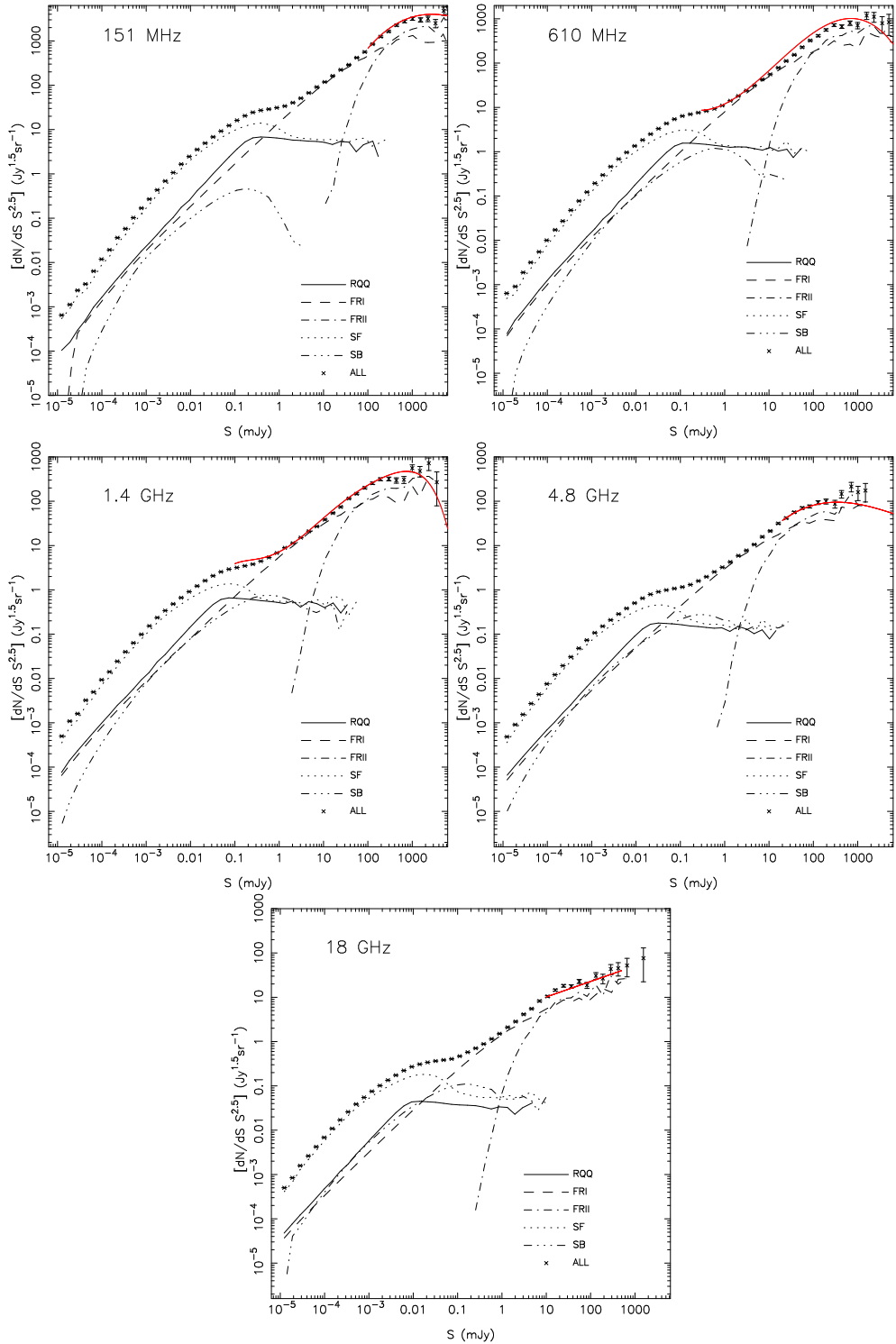

\includegraphics[width=6.5cm,angle=0]{scounts1.eps}
\includegraphics[width=6.5cm,angle=0]{scounts2.eps}
\includegraphics[width=6.5cm,angle=0]{scounts3.eps}
\includegraphics[width=6.5cm,angle=0]{scounts4.eps}
\includegraphics[width=6.5cm,angle=0]{scounts5.eps}
\caption{\normalsize Source counts generated from the simulation catalogue at the five frequencies, broken down according 
to source type. The red lines are polynomial fits to the observed source counts from the following papers: 151~MHz, Willott et al.~(2001); 610~MHz, Garn et al.~(2008); 1.4~GHz, Hopkins et al.~(2003); 4.86~GHz, Gregory et al.~(1994); in place of 18~GHz data we plot the 9C survey 15GHz source counts from Waldram et al.~(2003). In all cases, the simulated source counts are within the
scatter of the observations.}
\label{fig:dsc}
\end{figure*}

Differential source counts have been generated at the five observational frequencies (151~MHz, 610~MHz, 1.4~GHz, 4.86~GHz and 18~GHz) and are shown in Fig.~\ref{fig:dsc}. In all cases, they are in good agreement with the existing observational data. We 
reiterate that the input luminosity functions for the simulations were inferred from observations at either 151~MHz or 1.4~GHz 
(depending on the galaxy type), so this agreement demonstrates that we are modelling the radio spectral energy distributions 
in a realistic fashion. For the radio-loud AGN, the beaming parameters were, however, derived in order to reproduce the source counts above $\sim 1$~mJy (section 2.5).

Concerning the faint end of the source counts which are currently unconstrained by observational measurements, our model 
begins to diverge from the Hopkins et al.~(2000)/Windhorst~(2003) model below 1\microJy~at 1.4~GHz and by 10~nJy our 
counts are approximately a factor of 10 below theirs. This is almost entirely due to the inclusion in their model of a 
`normal galaxy' population derived from an optical luminosity function, supplementing the warm IRAS luminosity population. 
As Hopkins et al. point out, it is questionable whether this additional population is needed as the normal galaxies may already be adequately represented by the tail of the IRAS luminosity function. As discussed in section 2.4, our model works on the latter assumption. There is, however, need for an additional population of dwarf/irregular galaxies to account for the 
bulk of the HI mass function in range $10^{7-8}$\Msun~(section 3.7), but such galaxies are expected to be fainter in the 
radio continuum (compared with normal galaxies) and it is unclear how to incorporate them in these continuum simulations.

\subsection{Local luminosity functions}
\begin{figure}
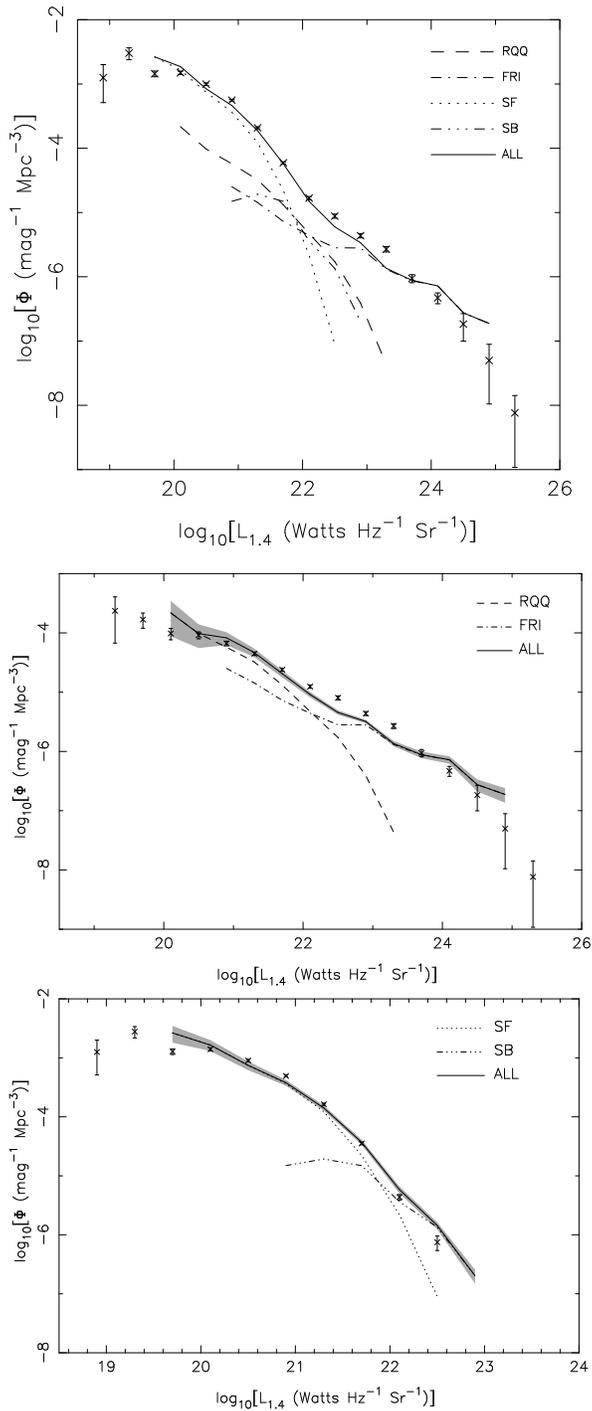

\includegraphics[width=7.5cm,angle=0]{local_rlf.eps}
\includegraphics[width=5.5cm,angle=270]{rlf_agn.eps}
\includegraphics[width=5.6cm,angle=270]{rlf_sf.eps}

\caption{\normalsize The uppermost plot shows the contributions of the various populations to the 
local ($z<0.3$) 1.4~GHz luminosity function in the simulation, for flux densities $>2.5$~mJy. 
The data points show the luminosity function derived for the same flux density limit for NVSS sources 
by Mauch \& Sadler~(2007). The middle and lower plots show the comparison for the AGN and 
star-forming galaxies separately, with the grey regions indicating the Poisson errors on the simulations.}
\label{fig:lfs}
\end{figure}

The local 1.4~GHz luminosity functions for the individual galaxy types and for the radio source population as a 
whole have been constructed and are shown in Fig.~\ref{fig:lfs}. `Local' is defined here as meaning $z<0.3$ and a 
lower flux density cut of 2.5~mJy at 1.4~GHz has been applied, in order to match the NVSS. The luminosity functions 
are in generally good agreement with those determined by Mauch \& Sadler~(2007) from a 6dF spectroscopic survey of NVSS radio sources. When the comparison is made for the AGN and star-forming galaxies separately, discrepancies appear at luminosities $22 < \rm{log} L_{\rm{1.4GHz}} < 23$ [W/Hz/sr], which may be due to the 
observational mis-classification of some of the AGN as star-forming galaxies, as discussed in section 2.4.

\subsection{Angular clustering}

As discussed in section 2, the incorporation of cosmological clustering in the galaxy distribution constitutes one of the
main innovations of our simulations. A simple test of this aspect is provided by the angular correlation function, $w(\theta)$, of 
radio sources, for which numerous measurements now exist from the FIRST and NVSS radio surveys at 1.4~GHz (see Overzier et al.~2003 
and Blake \& Wall~2002a,b). As discovered by the latter papers, the observed $w(\theta)$ can be described by a double power-law: 
(i) a steep power law of the form $w(\theta) = A \theta^{-3.4}$, dominant below 0.1~degrees, which is due to substructure (i.e. 
cores, lobes and hotspots) within individual radio galaxies; and (ii) a cosmological signal of the form 
$w(\theta) = B \theta^{-0.8}$ which dominates on larger scales. According to Overzier et al.~(2003), the amplitude of the 
cosmological clustering measured in NVSS/FIRST is essentially constant from 1.4~GHz flux density limits from 3-30~mJy, and then 
appears to increase by a factor of 10 above 200~mJy. However, a re-analysis of angular clustering in the NVSS, WENSS and SUMSS 
surveys by Blake et al.~(2004) found no evidence that the amplitude depends on the 1.4~GHz flux density over the range 10--200~mJy,
although the associated error bars are large. The $w(\theta)$ measurements at these flux limits probe the clustering of just the 
powerful FRI/FRII sources and it would be necessary to extend them well below 1mJy (or even 100$\mu$Jy) in order to discern the 
impact of the star-forming galaxies on the clustering signal in an unidentified radio survey (Wilman et al.~2003). 

\begin{figure}
\includegraphics[width=7.5cm,angle=0]{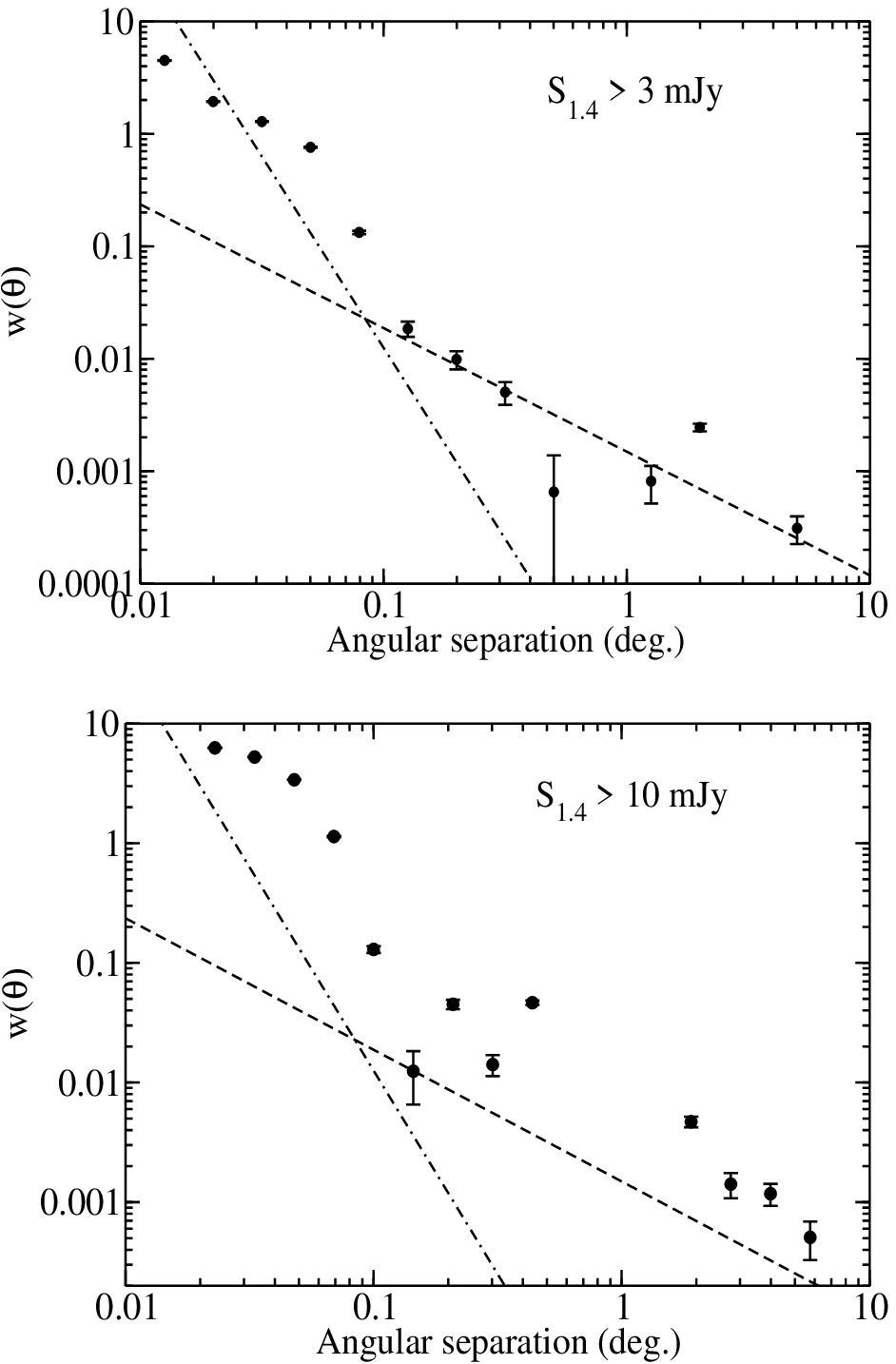}
\caption{\normalsize The angular 2-point correlation function, $w(\theta)$, measured in our simulations for 1.4~GHz flux limits of 3mJy and 10mJy. The lines overplotted in both cases are the cosmological (dashed line) and multi-component (dot-dashed line) contributions to $w(\theta)$ measured by Blake et al.~2004 for a 10mJy flux limit. The error bars are Poisson error bars, which are not strictly correct for clustered data.}
\label{fig:wtheta}
\end{figure}

Our $w(\theta)$ measurements are shown in Fig.~\ref{fig:wtheta} for 1.4~GHz flux limits of 3 and 10~mJy. Qualitatively, the
simulations reproduce the observed double component structure in $w(\theta)$ with the break in the observed position (0.1 degrees) 
and the correct slope and amplitude for the cosmological component. There are some deviations from a power-law form in the
multi-component contribution to $w(\theta)$ but these just reflect our simplistic modelling of the internal structures of 
individual radio sources. The amplitude of the simulated cosmological clustering signal at 3mJy is in agreement with the `base 
level' amplitude measured by Overzier et al.~(2003) at 3--30~mJy. At higher flux densities, however, the simulated 
amplitude appears to increase rather faster than observed. This is likely to be the result of the relatively crude bias prescription in the simulations, in which a single halo mass is assigned to each population, independent of luminosity and redshift: in reality, there may be be a smooth transition in halo mass from the FRIs (for which we assumed $10^{13}h^{-1}$\Msun)~to the 
FRIIs (modelled with $10^{14}h^{-1}$\Msun).

\subsection{Spectral index distributions}
\begin{figure}
\includegraphics[width=6.5cm,angle=0]{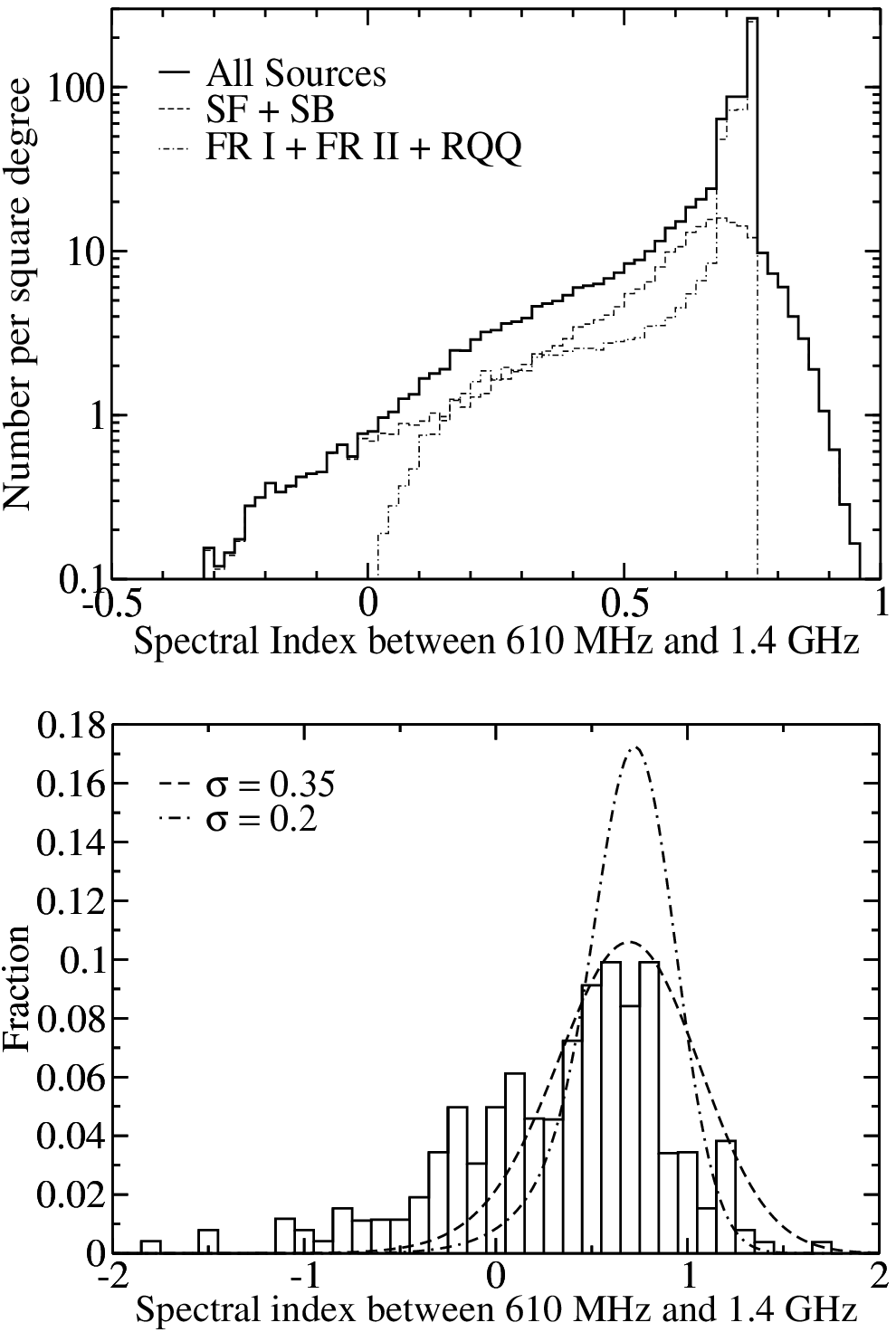}
\caption{\normalsize The upper plot shows the observed 610~MHz--1.4~GHz spectral index distribution for our simulation, broken
down into separate populations of AGN and star-forming galaxies (the spike around 0.7--0.8 is due to the radio-quiet AGN and 
the extended emission of the radio-loud AGN, which are modelled without scatter). The lower plot shows an observed distribution derived by Garn et al.~(2008) and the effect of gaussian smoothing the simulated distribution, in an attempt to better match the observations.}
\label{fig:sids}
\end{figure}

The distributions of observed radio spectral index between 610~MHz and 1.4~GHz are shown in Fig.~\ref{fig:sids}, broken
down into the separate distributions for the star-forming galaxies and the AGN (spectral index $\alpha$ is here defined with
the convention $S_{\rm{\nu}} \propto \nu^{-\alpha}$). Flux density limits of $S_{\rm{610 MHz}} > 360\mu$Jy and $S_{\rm{1.4 GHz}} > 200\mu$Jy~were applied to match the selection criteria of Garn et al.~(2008), whose spectral index
distribution is shown for comparison. There are clearly discrepancies between the two, particularly the failure of the
simulations to reproduce the flat spectrum tail of the observed distribution, which might be partially due to measurement 
errors at the flux density limit in the observations and also in part due to an excessive amount of high-frequency curvature 
in the assumed spectra of the radio-loud AGN cores (as defined in section 2.5). Fig.~7 also shows the effect of 
gaussian-smoothing the simulated spectral index distribution during post-processing.

\subsection{P-D and D-z diagrams}
\begin{figure}
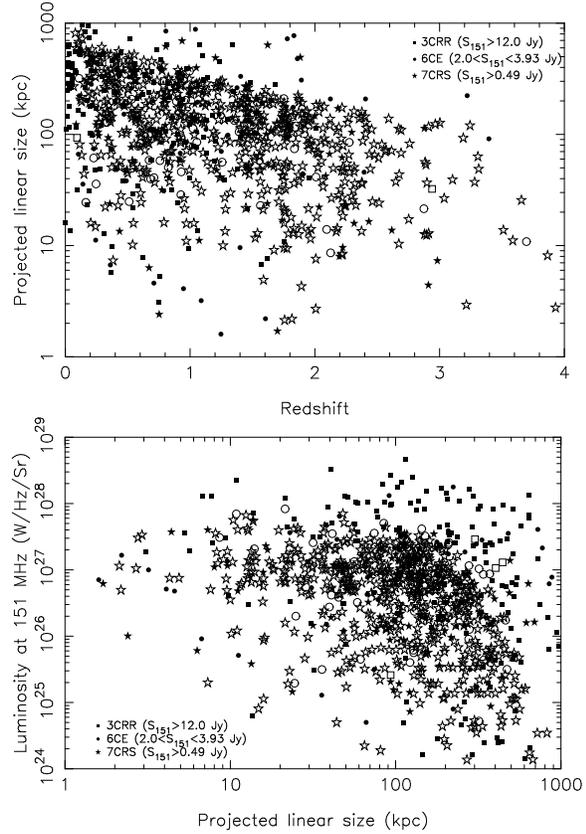

\includegraphics[width=5.5cm,angle=270]{p-z-large.eps}
\includegraphics[width=5.5cm,angle=270]{p-d-large.eps}
\caption{\normalsize A comparison of the projected linear size-redshift and radio power- projected linear size relations of the simulated sources (open points) with those in the 3CRR, 6CE and 7CRS surveys (filled points). Flux density cuts have been applied to the simulated sources to mimic these surveys. See text for discussion.}
\label{fig:pdz}
\end{figure}

In Fig.~\ref{fig:pdz} we compare the 151~MHz radio luminosity-redshift and projected linear size-redshift relations of the simulated sources
with observational data from the 3CRR, 6CE and 7CRS surveys (e.g. Blundell, Rawlings \& Willott~1999; BRW). Flux density cuts have
been applied to the simulated sources to match these surveys. Despite the simplicity of the algorithm for applying true linear 
sizes to our simulated sources (section 2.5), there is good agreement between the projected linear sizes of the simulated 
powerful radio sources and those in real low-frequency-selected redshift surveys. In both plots of Fig.~\ref{fig:pdz},
the main differences between the simulated and real
data points are readily explicable as being due either: (i) to the differences in area between the
semi-empirical simulation and the real datasets; or (ii) to known inadequacies in the simulation
technique. Examples of case-(i) differences include: the semi-empirical survey has a sky area
a factor $\simeq 35$ smaller than 3CRR (0.12 vs 4.239~\sr), explaining why no simulated sources appear in the
upper dex of radio power (i.e. $28 \leq \log L_{\rm{151}} < 29$ [W/Hz/sr]) in Fig.~\ref{fig:pdz}, compared with
37 3CRR sources; the small number of real objects beyond $z \sim 3$ in Fig.~\ref{fig:pdz} (to which 3CRR is insensitive) 
reflects the fact that the simulation covers 5.5 times the area of the 7CRS (0.12 vs 0.022~\sr). An example of a 
case-(ii) difference is that the simulation hard-wires a $D_{\rm{true}}=1$\Mpc~cut-off in size, which declines with 
redshift, whereas a few larger (giant) sources larger than this redshift-dependent limit
are seen in the real data out to $z \sim 2$. There is clearly much scope for ameliorating such problems by 
applying more sophisticated radio sources evolution models, although as emphasised by BRW this 
needs to be done in a self-consistent way that simultaneously models
the space-and-time-dependent spectral indices of the extended radio-emitting
components and also models variations in the radio source environments.

\subsection{HI mass function}
The main focus of these simulations is the radio continuum emission. As discussed in section 2.6, however, we can use the loose
correlation between HI mass and star formation rate to assign HI masses to the star-forming galaxies. Although this relation was 
measured for the `normal galaxy' population, we also apply it to the starburst galaxies, even though it is unlikely to hold for 
the most extreme objects. We did not attempt to assign HI masses to the AGN populations because there are no simple 
observationally-based prescriptions for doing so. Moreover, the AGN are most likely to reside in early-type galaxies, whose
contribution to the local HI mass function is significantly less than that of the late-types (Zwaan et al.~2003).

\begin{figure}
\includegraphics[width=7.5cm,angle=0]{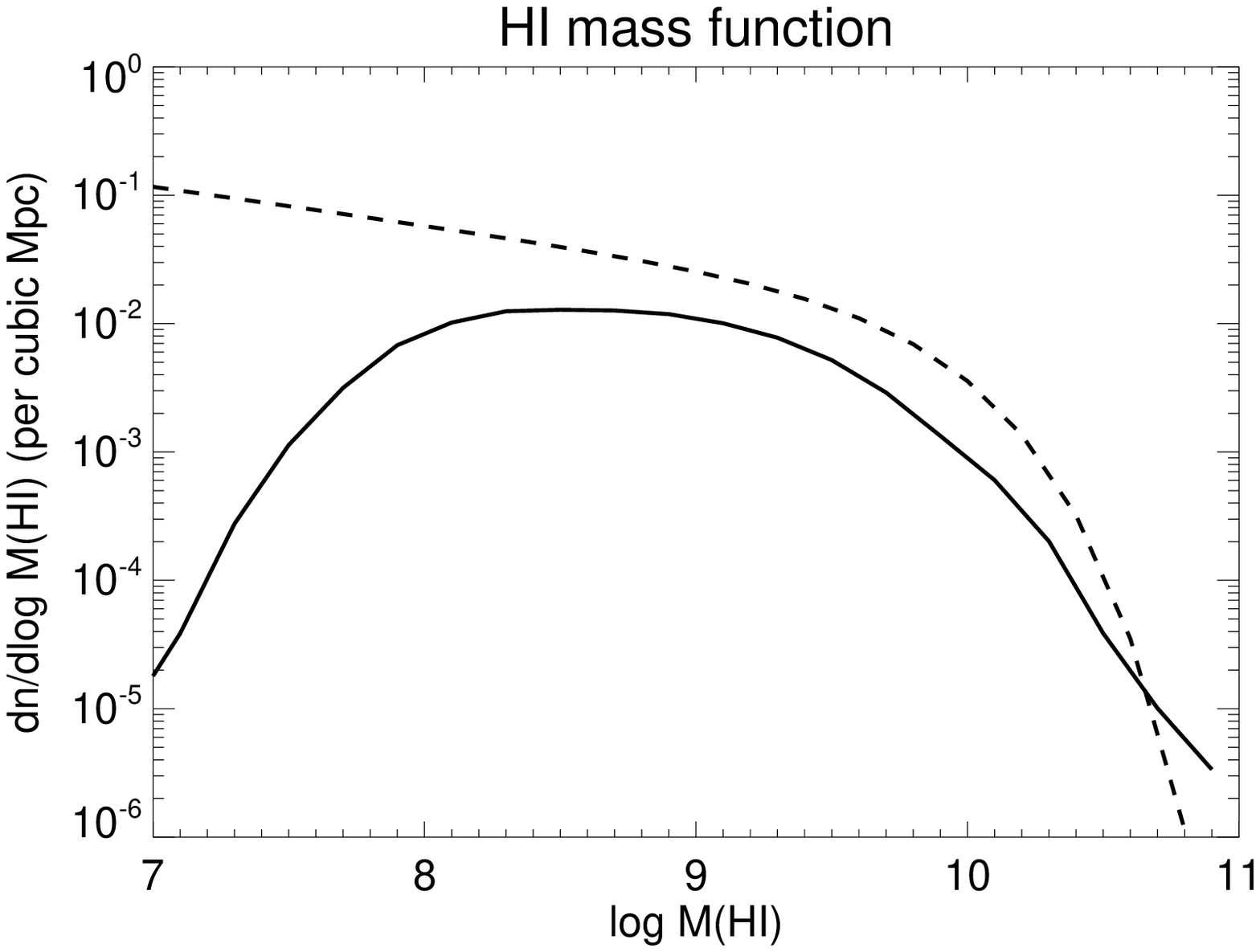}
\caption{\normalsize The solid line shows the HI mass function derived from the star-forming galaxies in the simulation over the
redshift range $z=0-0.1$. The dashed line is the local HIPASS mass function from Zwaan et al.~(2003).}
\label{fig:HImassfn}
\end{figure}

The $z=0-0.1$ HI mass function of the star-forming galaxies in the simulation is shown in Fig~\ref{fig:HImassfn}, 
where it is compared with that from the HI Parkes All Sky Survey (HIPASS) (Zwaan et al.~2003). The excess contribution in 
the simulations over the HIPASS fit above $10^{10.7}$\Msun~is due to the contribution of the starburst galaxies which, as 
discussed above, are likely to have lower HI masses than those implied by the HI mass--star formation rate relation of 
the normal galaxies. The divergence between the simulations and the HIPASS fit at lower masses is chiefly due to the absence in our simulation of a dwarf/irregular galaxy population; Zwaan et al. showed that this population (morphological
type ``Sm-Irr'') accounts for an increasing proportion of the HI mass function as the mass is reduced below $10^{10}$\Msun, and
essentially for all of it in the range $10^{7-8}$\Msun. However, such a population is probably not accounted for in 
our normal galaxy radio luminosity function, even though the lower integration limit for the latter of 
log $L_{\rm{1.4}} = 17$~[W/Hz/sr] equates to a very modest star formation rate of around $10^{-3}$\Msunpyr. Compared with normal 
galaxies, dwarf galaxies are gas-rich objects, with higher ratios of HI mass to star formation rate 
(see e.g. Roberts \& Haynes~1992), possibly because conditions in the dwarf galaxies do not meet disc instability criteria necessary for the onset of star formation.

A much fuller treatment of HI is provided by the SKADS semi-analytic simulations of Obreschkow et al.~(2008).

\section{DEFICIENCIES OF THE SIMULATION}
Here we reiterate what we regard as the most important 
weaknesses in our modelling approach that users should be aware of. Some of these issues can be circumvented with appropriate
post-processing, whereas others are hard-wired into the simulation and therefore of a more fundamental nature. \\

\noindent
$\bullet$ {\em The use of extrapolated luminosity functions.} The extrapolation of luminosity functions beyond the regimes of
luminosity and redshift in which they were determined is unavoidable when attempting to simulate a next generation facility 
with a quantum leap in sensitivity over existing telescopes. The most important aspect of this concerns the faint 
end of the normal galaxy radio luminosity function, which is controversial even in the regime already observed. As discussed
in section 2.4, we assume that the luminosity function flattens below log L$_{\rm{1.4 GHz}} = 19.6$~[W/Hz/sr], as determined
by Mauch \& Sadler~(2007), whilst others (e.g. Hopkins et al.~2000) have supplemented this with an additional population
derived from an optical luminosity function. We also assumed that the luminosity function of the star-forming galaxies does not evolve further in redshift from $z=1.5$ out to the redshift limit, $z=20$. This is clearly unrealistic, but it gives the user full freedom to implement any particular form of decline in the 
star formation rate as a post processing task, as multi-wavelength constraints on high-redshift star formation accrue in the 
coming years. For example, stronger forms of high-redshift decline in the space density can be implemented by random sampling the existing catalogue. For all luminosity functions, we have indicated the default post-processing option for negative high-redshift
evolution, and on the web database users will have the choice between implementing this or some other form.  \\

\noindent
$\bullet$ {\em The lack of star-forming/AGN hybrid galaxies and double counting.} The use of separate luminosity functions
for the AGN and star-forming galaxies is a fundamental design limitation of this simulation, and it prevents us from explicitly 
modelling hybrid galaxies where both processes contribute to the radio emission. However, such galaxies are implicitly present
in the simulation for the reason that at the faint end of the AGN radio luminosity function, star formation may make a non-negligible contribution to the radio emission. Similarly, some of the powerful starbursts may contain sizeable AGN contributions. These
effects will lead to a certain amount of double counting, perhaps over-producing the faint source counts. \\


\noindent
$\bullet$ {\em The lack of small-scale non-linear clustering}. Due to the use of $5 h^{-1}$\Mpc~cells and a linear-theory 
power spectrum, this simulation does not model small-scale non-linear clustering in a satisfactory manner. This is not 
considered a serious problem because the simulation was primarily designed to model the clustering on large-scales, which is
adequately described by our linear theory prescriptions. The `semi-analytical' simulations of
Obreschkow et al. (2008), based on N-body simulations, provide a much better description of small-scale non-linear
clustering. \\

\noindent
$\bullet$ {\em The treatment of galaxy clusters.} Clusters were identified in the mass density field using a numerical Press-Schechter filtering method. The coarse sampling of the density field and the sharpness of the real-space mass filter result in a 
quantisation of the cluster masses. Users with particular interest in the clusters would thus need to post-process the masses 
to obtain a continuous distribution to match the Sheth-Tormen~(1999) mass function (which provides a better fit to N-body 
simulations at the high mass end than the standard Press-Schecther formalism). It should also be noted that galaxies were 
assigned to the clusters in a very simplistic fashion, with no account of infall regions, for example. A more fundamental design
limitation of the simulation is that the underlying galaxy luminosity functions are assumed to be the same in the cluster and
field environments.

\section{CONCLUSIONS}
The semi-empirical simulation described in this paper has been designed to be used in an interactive fashion for optimising the design of the SKA
and its observing programmes in order to achieve its scientific goals. For this reason, we have made the galaxy catalogues 
available via a web-based database (http://s-cubed.physics.ox.ac.uk), which users can interactively query to generate 
`idealised radio skies' according to their requirements. Such skies can then be fed to (software) telescope simulators to facilitate the development of data processing and calibration pipeline routines. The extent to which the scientific 
content of the simulations can then be recovered at the end of such a simulation chain will enable observing programmes to be 
optimised for various models of the telescope hardware. In parallel with such work, we also anticipate that the simulations will be 
a valuable tool in the interpretation of existing radio surveys with abundant follow-up data.

\section*{ACKNOWLEDGMENTS}
RJW, HRK and FL are supported by the Square Kilometer Array Design Study (SKADS), financed by the European Commission. MJJ acknowledges support from SKADS and a Research Councils UK Fellowship. FBA acknowledges a Leverhulme Early Career Fellowship. We thank Alejo Martinez-Sansigre for discussions, and Anne Trefethen (Director of the OeRC) for the use of OeRC resources for the online database.

{}

\end{document}